\documentclass[%
aps,prl,preprint,amsmath,nofootinbib,amssymb
]{revtex4-2}

\def\vr{\textbf{r}}

\def\xc{\text{XC}}
\def\Exc{E_{\rm XC}}

\def\Ts{T_{\rm s}}
\def\EH{E_{\rm H}}
\def\vH{v_{\rm H}}

\def\vp{v_{\rm p}}
\def\Ep{E_{\rm p}}
\def\Ef{E_{\rm f}}
\def\nf{n_{\rm f}}
\def\na{n_{\rm \alpha}}

\def\Ea{E_{\rm \alpha}}
\def\Nf{N_{\rm frag}}
\def\sumNf{\sum_\alpha^{\Nf}}
\def\nad{{\rm nad}}

\def\bea{\begin{eqnarray}}
\def\eea{\end{eqnarray}}
\def\ben{\begin{equation}}
\def\een{\end{equation}}
\def\benu{\begin{enumerate}}
\def\enu{\end{enumerate}}

\def\n{n}
\def\lsim {\ifmmode {\buildrel<\over\sim}}

\def\sss{\scriptscriptstyle\rm}

\def\1var{(\bx_1...\bx\N)}

\def\bn{{\bf n}}
\def\br{{\bf r}}

\def\b1{{\bf 1}}
\def\bx{{x}}

\def\xc{_{\sss XC}}

\def\N{_{\sss N}}
\def\H{_{\sss H}}

\def\sph_int{ {\int d^3 r}}

\def\laplace1d{\frac{d^2}{dx^2}}
\def\plaplace1d{\frac{d^2}{d{x'}^2}}

\def\padr2{\frac{\partial^2}{\partial r^2}}

\def\N{{\cal N}}
\def\a{{\alpha}}
\def\b{{\beta}}

\def\Ts{T_{\rm s}}
\def\EH{E_{\rm H}}
\def\Exc{E_{\rm xc}}

\def\vp{v_{\rm p}}
\def\Ep{E_{\rm p}}
\def\Ef{E_{\rm f}}
\def\nf{n_{\rm f}}
\def\na{n_{\rm \alpha}}

\def\Ea{E_{\rm \alpha}}
\def\Nf{N_{\rm frag}}
\def\sumNf{\sum_\alpha^{\Nf}}
\def\nad{{\rm nad}}

\usepackage{graphicx}
\usepackage{dcolumn}
\usepackage{bm}       
\usepackage{mathptmx}
\usepackage[T1]{fontenc}

\begin{document}

\title{Stretching bonds in Density Functional Theory without artificial symmetry breaking} 
\author{Yuming Shi}
\affiliation{Department of Physics and Astronomy, Purdue University, West Lafayette, Indiana 47907, USA}
\author{Yi Shi}
\affiliation{Department of Chemistry, Purdue University, West Lafayette, Indiana 47907, USA}
\author{Adam Wasserman}
\email[]{awasser@purdue.edu}
\affiliation{Department of Physics and Astronomy, Purdue University, West Lafayette, Indiana 47907, USA}
\affiliation{Department of Chemistry, Purdue University, West Lafayette, Indiana 47907, USA}

\date{\today}

\begin{abstract}
Accurate first-principles calculations for the energies, charge distributions, and spin symmetries of many-electron systems are essential to understand and predict the electronic and structural properties of molecules and materials. Kohn-Sham density functional theory (KS-DFT) stands out among electronic-structure methods due to its balance of accuracy and computational efficiency. It is now extensively used in fields ranging from  materials engineering to rational drug design. However, to achieve chemically accurate energies, standard density functional approximations in KS-DFT often need to break underlying symmetries, a long-standing ``symmetry dilemma".  
By employing {\em fragment} spin densities as the main variables in calculations (rather than total molecular densities as in KS-DFT), we present an embedding framework in which this symmetry dilemma is resolved for the case of stretched molecules. The spatial overlap between fragment densities is used as the main ingredient to construct a simple, physically-motivated approximation to a universal functional of the fragment densities. This `overlap approximation' is shown to significantly improve semi-local KS-DFT binding energies of molecules without artificial symmetry breaking. 
\end{abstract}

\maketitle

\newpage

\begin{figure}[t]
\includegraphics[width=0.5\textwidth]{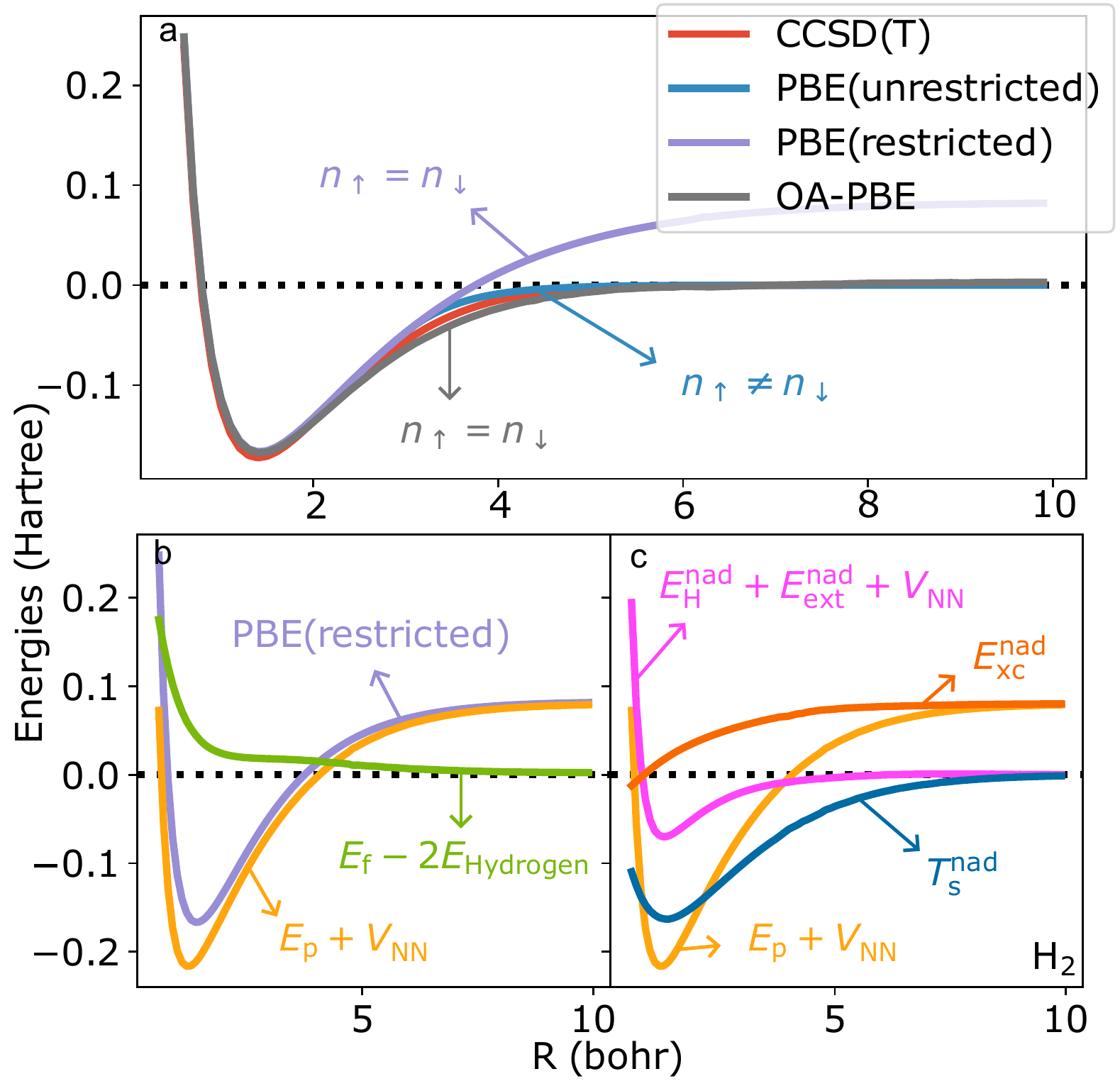}
\caption{\textbf{a}: Binding energy of H$_2$ calculated through: (i) CCSD(T) reference values (red); (ii) spin-unrestricted PBE (blue); (iii) spin-restricted PBE (purple); and (iv) OA-PBE from Eq. (\ref{e:OA-PBE}-\ref{e:overlap}) (gray).  \textbf{b}: restricted-PBE energies (purple) decomposed into: Fragment relaxation energies $\Ef-2E_{\rm Hydroden}$ (green) and partition energy $\Ep+V_{\rm NN}$ (orange). 
\textbf{c}: Decomposition of $\Ep$ (orange) into non-additive kinetic (blue), exchange-correlation (red) and all remaining contributions (pink).   Note that the large error of the restricted-PBE calculation as $R\to\infty$ can be attributed almost entirely to $E\xc^{\rm nad}$.
}
\label{fig:H2}
\end{figure}

Symmetry breaking can occur in the quantum-mechanical simulation of molecules when the lowest-energy solutions of the 
electronic Schr\"odinger equation (SE) do not exhibit the same symmetries of the underlying Hamiltonian. 
An exact solution of the SE or, equivalently, a solution of the Kohn-Sham equations of Density Functional Theory (KS-DFT) \cite{hohenberg1964inhomogeneous, kohn1965self} with the {\em exact} exchange-correlation (XC) functional $E\xc[n_\uparrow,n_\downarrow]$ yields spin-densities $\{n_\uparrow,n_\downarrow \}$ that retain the symmetry of the molecular Hamiltonian. However, it is well known that spin and charge symmetries of stretched molecules are often broken by {\em approximate} density-functional approximations (DFAs) for XC. Attempts to prevent such symmetry breaking often lead to qualitatively incorrect electric and magnetic properties of molecules and materials as a consequence of delocalization and static-correlation errors \cite{burke2012perspective, cohen2012challenges, mardirossian2017thirty, becke2014perspective, PPLB82, mori2009discontinuous}. 
Symmetry breaking, when allowed, can provide insight into the quantum-mechanical correlations 
that exist between fluctuating charges or spins in the constituent fragments. When these fragments are separated by a large distance $R\to\infty$, the correct (symmetry unbroken) solution of the SE is an infinite-time average over fluctuations among the possible broken-symmetry solutions. 

Consider for example the spin symmetry of a stretched hydrogen molecule in its singlet ground state. The correct spin-up density $n_\uparrow (\br)$  equals the spin-down density $n_\downarrow (\br)$ at every point in space, but imposing this symmetry on the solution of the KS-DFT equations with an approximate XC functional (see `restricted' in panel \textbf{a} of Fig. \ref{fig:H2} for the popular PBE \cite{PBE96}) leads to unacceptably large energy errors as the molecule is stretched beyond $R\sim 3$ bohr. 
A broken-symmetry solution exists with an energy that runs close to the exact one (unrestricted PBE in panel \textbf{a} of Fig. \ref{fig:H2}), with $n_\uparrow(\br)$ localized on one atom and $n_\downarrow(\br)$ on the other. Although strictly incorrect, this set of spin-densities does reflect one of the two possible dissociation channels observed when infinitesimal environmental perturbations induce the collapse of the wavefunction that breaks the chemical bond \cite{levine2020clarifying}. 
 
Similarly, recent studies show that the SCAN meta-GGA funcional \cite{SCAN15} can yield highly accurate binding energies for a great number of systems including some of the `strongly-correlated' type \cite{furness2018accurate, zhang2020competing, kitchaev2016energetics, liu2021spin, perdew2022symmetry}, but only when spin-symmetry breaking is allowed. 
A question then arises on the interpretation of such broken-symmetry solutions for {\em finite} $R$ \cite{perdew2021interpretations}. These are useful, among others,  to calculate values of magnetic properties of molecules such as exchange-coupling constants \cite{noodleman1981valence,dai2001spin}. 

Is it possible to calculate accurate energies {\em without} symmetry breaking when employing standard (e.g. PBE, SCAN) XC functionals? In this work we provide a positive answer. The key is to use a formulation of DFT in which: (1) Electronic {\em fragment} spin-densities (as an alternative to total molecular densities) are sharply defined for finite $R$ and recover those of isolated atoms as $R\to\infty$; (2) each of those fragment spin-densities is described through a mixed-state ensemble that can place fractional charges and spins on the fragments to guarantee the correct symmetries; and (3) there exists a universal functional of the set of fragment spin-densities that describes the fragment interaction and that is amenable to simple yet accurate approximations.  

\vspace{0.5cm}
{\bf {\em Strategy, results, and discussion:}} 
All three of these features are provided by Partition-DFT (P-DFT) \cite{elliott2010partition,nafziger2014density}, a density embedding method in which a molecule, defined by a nuclear ``external" potential $v(\br)=\sumNf v_\a(\br)$ and $N$ electrons,  is partitioned into $\Nf$ smaller fragments (labeled by $\alpha$). The features listed above are met in the following way: (1) The fragment spin-densities are uniquely defined by the requirement that the sum of fragment energies $\Ef$ be minimized under the constraint that the sum of fragment spin-densities $\n_{\rm{f}, \sigma}(\br)\equiv\sumNf n_{\alpha,\sigma}(\br)$ matches the correct spin-density $n_{\sigma}(\br)$ of the molecule, i.e. the ground-state spin-density for $N$ electrons in $v(\br)$. The Lagrange multiplier that enforces this constraint is a unique {\em partition potential} $\vp(\br)$ \cite{cohen2006hardness} ; (2) Each of the fragment spin-densities $n_{\alpha,\sigma}(\br)$ is a ground-state  {\em ensemble} density for a (possibly fractional) number of electrons and spins in $v_\a(\br)+\vp(\br)$; (3) A universal functional $Q[\bn]$ of the set of fragment spin-densities $\bn\equiv\{n_{\a,\sigma}\}$ is defined as: 
\ben 
Q[\bn]=F[n]-\sum_\a^{\Nf} \mathop{\rm min}_{\hat{\rho}_\a \to \{n_{\a,\uparrow},n_{\a,\downarrow}\}} Tr(\hat{\rho}_{\a}(\hat{T} + \hat{V}_{\rm ee}))~~,
\label{eq:Q}
\een  
where $F[n]=\mathop{\rm min}_{\Psi\to \{n_{\uparrow},n_{\downarrow}\}}\langle{\Psi|\hat{T} + \hat{V}_{\rm ee}|\Psi\rangle}$ is the Levy-Lieb functional of the total density \cite{levy1979universal, lieb1983density} , $\hat{T}$ and $\hat{V}_{\rm ee}$ are the kinetic and electron-electron repulsion operators, and the search inside the sum is performed over fragment density matrices $\hat{\rho}_\a$ yielding the preset pairs of fragment spin-densities $\{n_{\a,\uparrow},n_{\a,\downarrow}\}$.
When evaluated at the unique set $\bn$ of fragment spin-densities minimizing $\Ef$, the ground-state energy of the molecule is then given by:
\ben 
E = \Ef[\bn]+\Ep[\bn],
\label{eq:decomposition}
\een 
where $\Ef$ is the fragment energy summation without the contribution from the partition potential ($\Ef = \sumNf\Ea$) and the {\em partition energy} $\Ep$ has been defined as the rest $\Ep[\bn]=Q[\bn]+\int d\br v(\br) \nf(\br) - \sum_{\a,\sigma}^{\Nf} \int d\br v_{\a}(\br) \n_{\a,\sigma}(\br)$. It can be proven \cite{nafziger2014density} that the partition potential $\vp(\br)$ is the functional derivative of $\Ep$ evaluated for a given set of fragment spin-densities $\bn$. 

It is useful to decompose  $Q[\bn]$ in terms of the usual Kohn-Sham density functional quantities as the sum of three non-additive terms: $Q[\bn]=\Ts^{\rm nad}[\bn]+E\H^{\rm nad}[\bn]+E\xc^{\rm nad}[\bn]$, where, e.g., $\Exc^\nad[\bn]=\Exc[\nf]-\sumNf\Exc[\na]$. One can then see that most of the PBE error for stretched H$_2$, for example, is contained in $E_p$ (panel {\bf b} of Fig. \ref{fig:H2}) and, more specifically, in $E\xc^{\rm nad}$ (panel $\bf c$). Various strategies for approximating the KS kinetic term, $\Ts^{\rm nad}[\bn]$, are being investigated \cite{shao2021gga, jiang2018non, polak2023symmetrized} but here we compute this term  {\em exactly} via density-to-potential inversions \cite{wu2003direct, shi2021inverse, shi2022n2v, kanungo2019exact}. Thus, for a given approximation to $E\xc[n]$, the P-DFT calculations simply reproduce the results of KS-DFT, including all of their errors (purple line in panel $\bf a$ of Fig. \ref{fig:H2}). 
In this article, we argue that almost all of the error of PBE at dissociation can be attributed to the incorrect behavior of $E\xc^{\rm nad}(R)$ and can be suppressed through improved approximations for this term alone. The gray line label OA-PBE in panel {\bf a} of Fig. \ref{fig:H2}, for example, shows how a simple `overlap approximation' for $E\xc^{\rm nad}[\bn]$ (to be defined below) removes most of the PBE error while preserving the correct spin symmetry as $R\to\infty$. 

\begin{figure}
\includegraphics[width=0.9\textwidth]{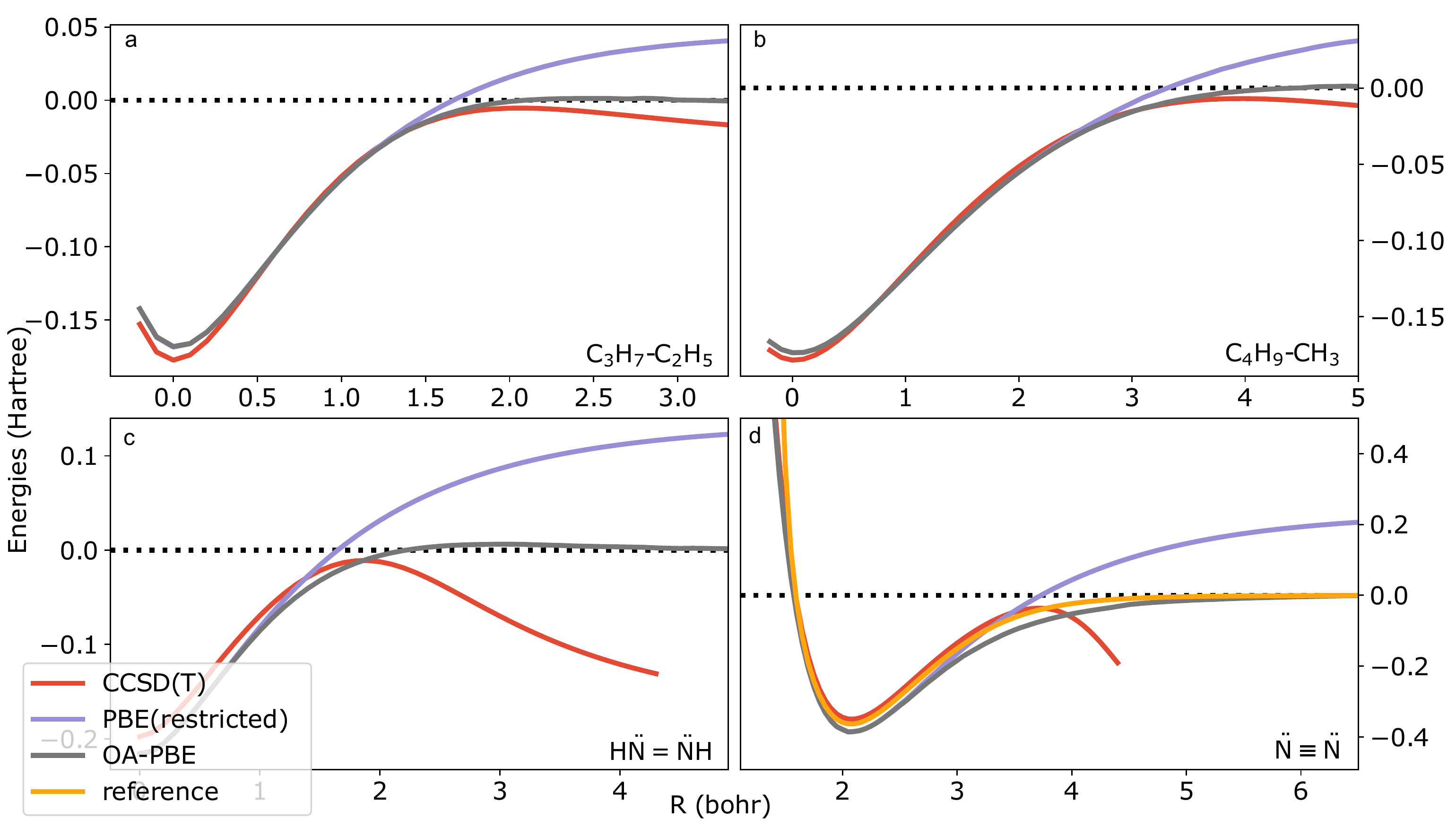}
\caption{Using PBE for the fragments, Eq. (\ref{e:OA-PBE}-\ref{e:overlap}) with $C=2$ and $p=1/2$ yield accurate binding energies (gray) when cutting through single ($N_b=1$), double ($N_b=2$), and triple ($N_b=3$) bonds. Comparisons are made with spin-restricted PBE (purple), CCSD(T) (red), or reference values from Ref. \cite{le2006accurate} (yellow).}
\label{fig:molecules}
\end{figure}

We demonstrate that accurate binding energies for stretched molecules can be obtained through physically-motivated approximations for $E\xc^{\rm nad}[\bn]$ {\em without} symmetry breaking.
We begin with the simplest case of closed-shell molecules partitioned into $\Nf=2$ fragments with spin-summed fragment densities $n_A(\br)$ and $n_B(\br)$ using a standard GGA funcional (PBE). For all such cases, like in H$_2$,  $E^{\rm nad}_{\rm xc, PBE}(R)$ goes to an incorrect positive constant as $R\to\infty$ rather than satisfying the {\em exact constraint}: $E\xc^{\rm nad}[\bn](R)\to 0$. We now build this constraint into $E\xc^{\rm nad}[\bn]$ through:
\begin{eqnarray}
E_{\rm xc, PBE}^{\rm nad, OA}[\bn]=S[\bn]E_{\rm xc, PBE}^{\rm nad},
\label{e:OA-PBE}
\\
S[\bn]={\rm erf}\left[\frac{C}{N_b}\int d\br\left(n_A(\br)n_B(\br)\right)^p\right],
\label{e:overlap}
\end{eqnarray}
where $N_b$ is the bond order.
When the two parameters $C$ and $p$ in Eq. (\ref{e:overlap}) are fixed as $C=2$ and $p=1/2$ (fitted for H$_2$), one obtains the binding curve labeled ``OA-PBE" in Fig. \ref{fig:H2}. 
Although $N_b$ in Eq. (\ref{e:overlap}) is itself a functional of the set of fragment densities, we find that the simple rule where $N_b=1$ for single, 2 for double, and 3 for triple bonds, is adequate for preserving the description of PBE around equilibrium while correcting the PBE errors as these build up beyond the Coulson-Fisher point. 
Eq. (\ref{e:OA-PBE}-\ref{e:overlap}) perform extremely well for other singly-bonded hydrocarbons ($N_b=1$) as well as doubly-bonded diazene ($N_b=2$) and triply-bonded nitrogen molecules ($N_b=3$), systems that are famously challenging for the standard DFAs in DFT \cite{cohen2012challenges}, and also for the `gold standard' of quantum chemistry methods, CCSD(T) (coupled-cluster with single and double and perturbative triple excitations) \cite{kowalski2000renormalized}, see Fig. \ref{fig:molecules}. 

\begin{figure}
\includegraphics[width=0.5\textwidth]{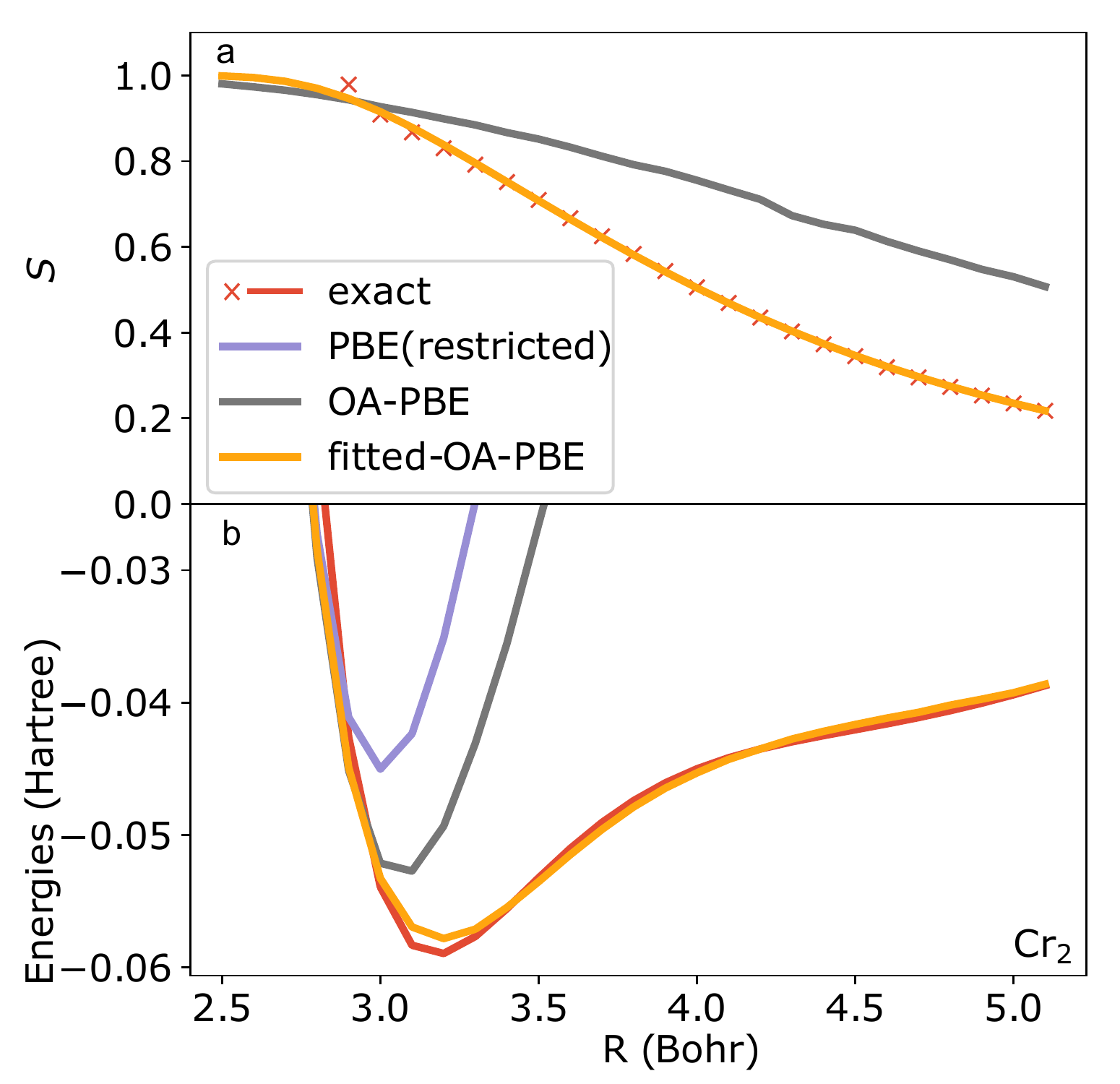}
\caption{\label{fig:Cr2} \textbf{a}: $S(R)$ for Cr$_2$ obtained from: (i) Numerically
exact results from ref. \cite{larsson2022chromium} (red, exact); (ii) Eq. (\ref{e:overlap}) with $p = 1/2$, and
$N_b = 6$ (gray, OA-PBE); and (iii) Fitted with $N_b=1.176$, $p=0.8$ (yellow, fitted-OA-PBE). \textbf{b}:
Corresponding binding energies, where pure restricted-PBE has been included for comparison (purple).
}
\end{figure}

The overlap functional as defined by Eq. (\ref{e:overlap}) is inadequate for molecules that are even more `strongly-correlated' than N$_2$. To illustrate this point, consider the challenging case of the chromium dimer (Cr$_2$). A quantitative description of the electronic structure of Cr$_2$ is a stringent test for any theory that attempts to capture strong correlations in molecules. Neither CASSCF nor CCSD(T) yield quantitative agreement for the ground-state energy of Cr$_2$ as a function of the inter-nuclear separation. As is well known, standard DFAs in KS-DFT are utterly inadequate to capture the multi-reference character of the ground state in stretched Cr$_2$. Only very recently has a truly {\em ab initio} calculation been reported for Cr$_2$ \cite{larsson2022chromium}. Unsurprisingly, Fig. \ref{fig:Cr2} shows that the OA-PBE of Eq. (\ref{e:OA-PBE})-(\ref{e:overlap}) with $N_b=6$ (gray line in the \textbf{b} panel) does improve (red) but not enough to match the {\em ab initio} results, as the inter-fragment interaction in Cr$_2$ is radically different than in the molecules of Fig. \ref{fig:molecules}. Nevertheless, the yellow line in the \textbf{b} panel of Fig. \ref{fig:Cr2} demonstrates that using  $p=0.8$ instead of $p=0.5$ in the integrand of Eq. (\ref{e:overlap}), and the value $N_b=1.176$ in the denominator, one obtains quantitative agreement between the OA-PBE and the most accurate (but expensive) state-of-the-art {\em ab initio} calculations for all inter-nuclear separations.
The obvious question is how the overlap functional should be defined for the OA-PBE to become more generally applicable and predictive. The simplicity of Eq. (\ref{e:OA-PBE}) and of the forms we have used for $S[\bn]$ suggest that it should be possible to derive a more general functional for $S[\bn]$ from first principles.

If one had access to the {\em exact} XC functional, Eq. (\ref{e:OA-PBE}) with ``PBE" replaced by ``exact" would evidently hold with $S=1$ for any $R$.  It is then useful to re-write Eq. (\ref{e:overlap}) as:
$S_{\rm exact}[\bn]=E_{\rm {xc,  exact}}^{\rm nad}[\bn] /E_{\rm {xc, DFA}}^{\rm nad}[\bn]$,
where $E_{\rm {xc, DFA}}^{\rm nad}[\bn]$ is the non-additive exchange correlation energy obtained through a self-consistent P-DFT calculation that uses a DFA for the XC energy functional. The overlap functional is thus approximating the missing (mostly non-local) effects in the DFA.  With access to accurate total energies, one can extract $E_{\xc, \rm exact}^{\rm nad}[\bn]$ by subtraction of the other components available from exact P-DFT calculations. The behavior of $S_{\rm exact}(R)$ can then be examined as illustrated for the case of Cr$_2$ in the \textbf{a} panel of Fig. \ref{fig:Cr2}, where the total electronic energies reported in ref. \cite{larsson2022chromium} were used as input. The smoothness of $S_{\rm exact}(R)$ (see \textbf{a} panel of Fig. \ref{fig:Cr2}), especially in regions  where $E(R)$ varies quite rapidly with $R$ (see \textbf{b} panel of Fig. \ref{fig:Cr2}) illustrates the usefulness of  Eq. (\ref{e:OA-PBE}-\ref{e:overlap}) for modeling the electronic structure  of a strongly-correlated molecule, an  indication that the path is worth pursuing further.

\begin{figure}[t]
\includegraphics[width=0.5\textwidth]{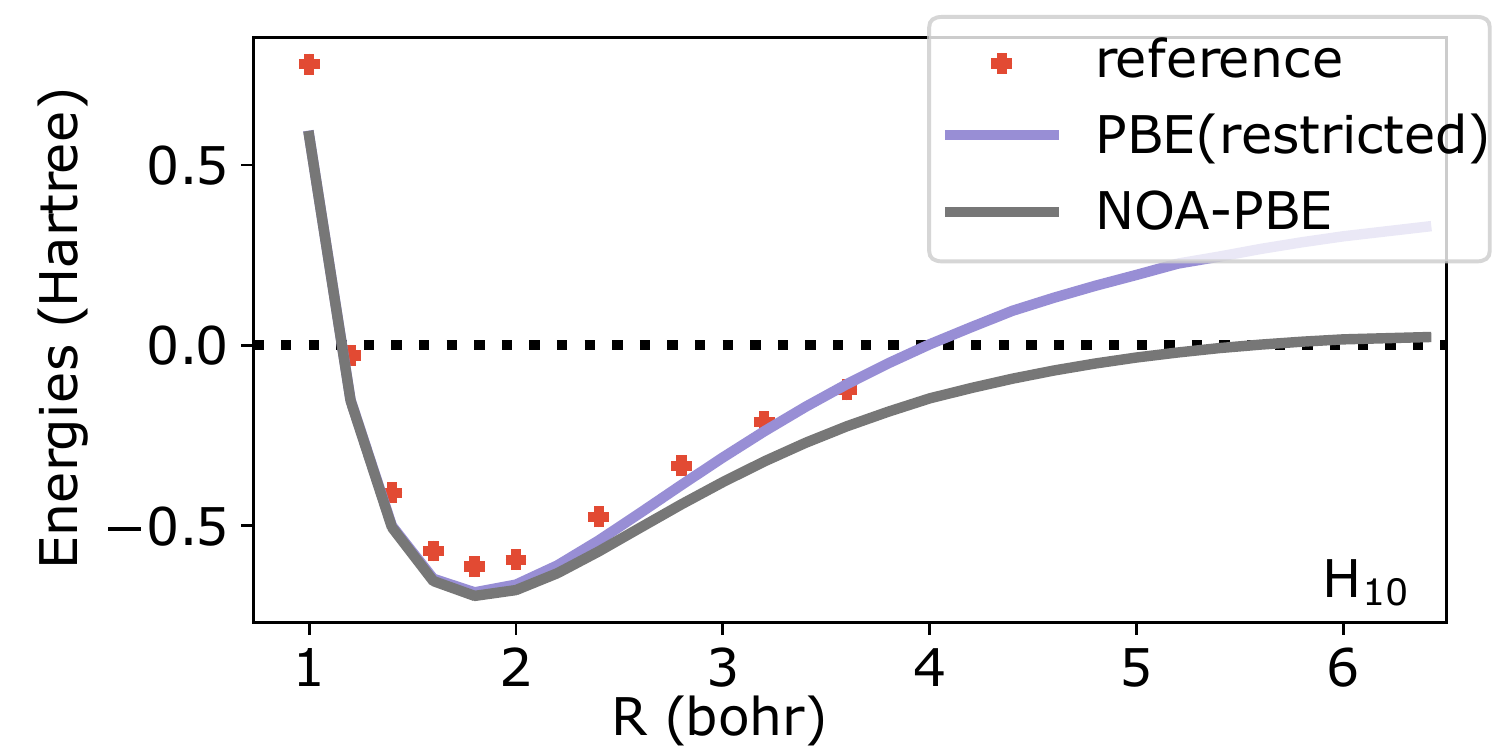}
\caption{\label{fig:H10} The NOA of Eq. (\ref{eqn:NOA}) corrects the PBE error for the binding energies in
hydrogen chains as $R\to\infty$ (here H$_{10}$, where
the x-axis is the distance $R$ between neighboring
nuclei). The reference is Multireference Configuration Interaction taken from Ref. \cite{motta2017towards}.
}
\end{figure}
Before moving on to considering the case of charge symmetry, we provide a proof-of-principle demonstration that the same idea of Eq. (\ref{e:OA-PBE}-\ref{e:overlap}) can be extended to an arbitrary number of fragments $\Nf>2$. When there are more fragments and P-DFT yields a set of densities $\bn=\{n_1(\br),n_2(\br),...,n_{\Nf}(\br)\}$, Eq. (\ref{e:OA-PBE}) can be applied recursively as a nested version of the OA (NOA):

\ben
E\xc^{\rm nad,NOA}[\bn]=S[\bn_{1\to m},\bn_{m+1\to \Nf}]E\xc^{\rm nad}[\bn_{1\to m},\bn_{m+1\to \Nf}]+E\xc^{\rm nad, NOA}[\bn_{1\to m}]+E\xc^{\rm nad, NOA}[\bn_{m+1\to \Nf}]
\label{eqn:NOA}
\een
where $\bn_{a\to d}$ denotes the partial sum of fragment densities $n_a(\br)+{...}+n_d(\br)$.
As for the case of binary fragmentation, this prescription preserves the results of the parent DFA at equilibrium separations and can improve the results when bonds are stretched. We have tested Eq. (\ref{eqn:NOA})  on hydrogen chains with the overlap model of Eq. (\ref{e:overlap}) and the results demonstrate that Eq. (\ref{eqn:NOA}) corrects the errors of PBE as $R\to\infty$ (see Fig. \ref{fig:H10} for H$_{10}$, a well known test-bed for strongly-correlated systems \cite{motta2017towards}), although it overestimates the corrections needed in the intermediate range $2.5<R<4$ bohr. More research into the form of $E\xc^{\rm nad}[\bn]$ and its accompanying overlap measure is clearly needed.

\begin{figure}[t]
\includegraphics[width=0.9\textwidth]{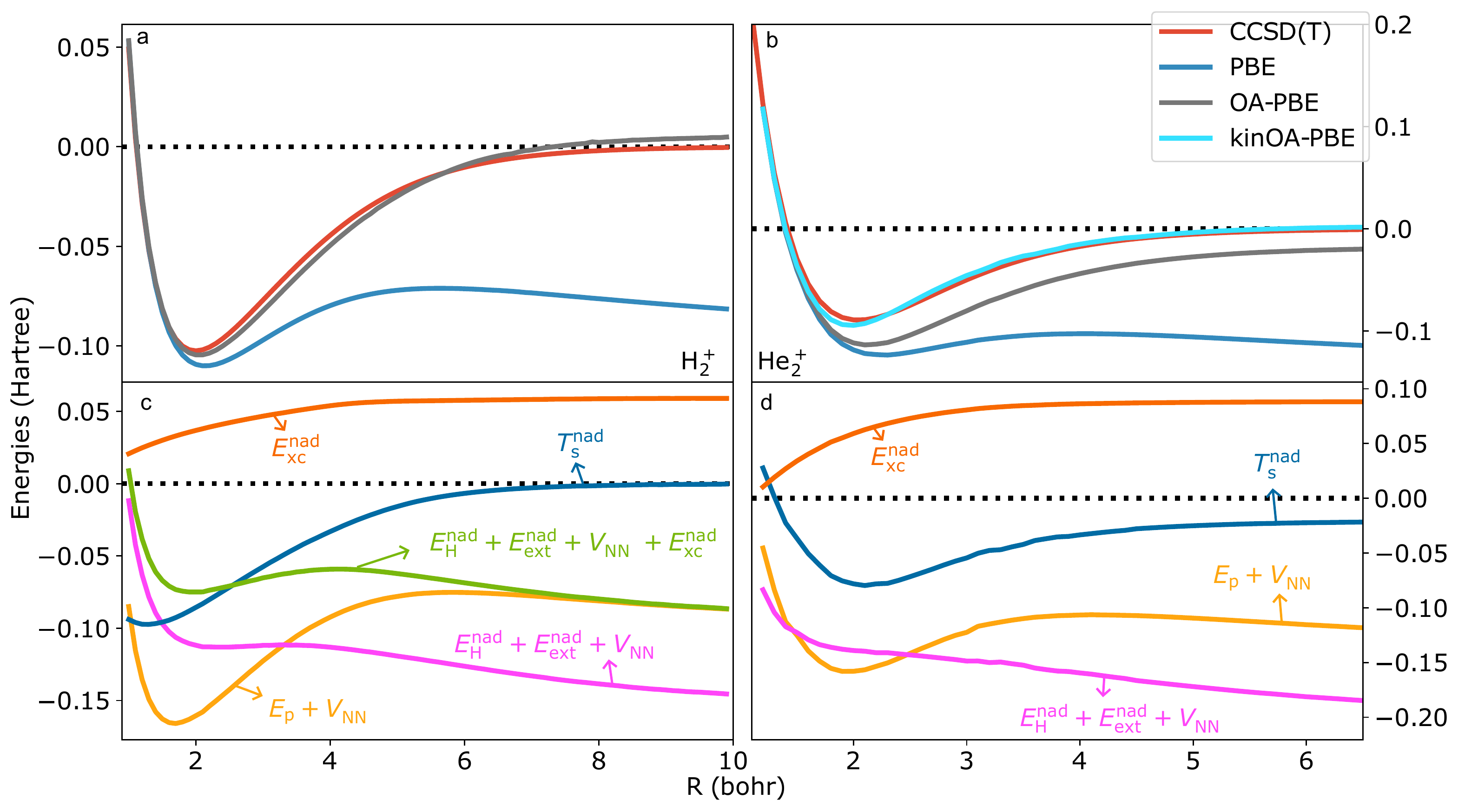}
\caption{\label{fig:H2p-and-He2p} \textbf{a}: H$_2^+$ binding energies calculated with CCSD(T) (red), PBE (blue), and OA-PBE (gray). \textbf{b}: He$_2^+$ binding energies, including a kinOA calculation in which $Q= S\Ts^{\rm nad}+E\H+E_{\rm xc, PBE}^{\rm nad,OA}[\bn]$ (light blue). \textbf{c,d}: Components of $\Ep$ showing that the PBE error can be attributed to a poor cancellation of errors between $E\xc^{\rm nad}$ and $E\H^{\rm nad}$.
}
\end{figure}

We now discuss {\em charge symmetry}, which is analogous to the case of spin symmetry but with an extra challenge. First, the analogy: As in the case of spin symmetry we have just discussed, the approximation chosen for $E\xc[n_{\uparrow},n_{\downarrow}]$ will typically lead to improved energies when charge symmetries are broken. The extra challenge: The charge-symmetry-broken solutions are typically {\em higher} in energy that the charge-symmetric solutions, and will therefore not be found when searching for a minimum. In other words: The analog of spin-unrestricted calculations will not lead to improved energies. Take for example the case of stretched H$_2^+$ in Fig. \ref{fig:H2p-and-He2p}, where PBE underestimates the dissociation energy by $70\%$. The PBE ground-state energy of an isolated hydrogen atom is only off by $0.08\%$,
so the dissociation energy error can be attributed almost entirely in this case to the fact that the KS equations do {\em not} break the charge symmetry of the ground state. How can one keep that symmetry {\em and} correct the energy? Again, as before, we analyze the contributions to $Q[\bn]$ from the different KS components and find that, this time, the problem is not fixed by simply quenching $E\xc^{\rm nad}(R)$ for large $R$ because, for any given $R$, $\Exc^{\rm nad}(R)$ does not cancel $\EH^{\rm nad}(R)$ as it should for a 1-electron system (see panel ${\bf c}$ of Fig. \ref{fig:H2p-and-He2p}). This {\em non-additive self-interaction error} can be corrected by adding a term to $\Exc^{\rm nad}$ in Eq. (\ref{e:OA-PBE}):
\begin{eqnarray}
E_{\rm xc}^{\rm nad, OA}[\bn]=S[\bn]E_{\rm xc, PBE}^{\rm nad}+(1-S[\bn])\Delta E\H^{\rm nad}
\label{e:OA-PBE-II}\\
\Delta\EH^\nad=g_{ij}\int\frac{n_{\mathrm{A}, i}(\vr)n_{\mathrm{B}, j}(\vr')}{|\vr - \vr'|}d\vr d\vr' - \EH^\nad,
\label{e:dEHnad}
\end{eqnarray}
where $g_{ij}=1$ when $N_{\rm A,\sigma}+N_{\rm B,\sigma}=N_{\sigma}$; and $g_{ij}=0$ otherwise, implying the possible dissociation channels. 
With this choice of $g_{ij}$, Eq. (\ref{e:OA-PBE-II}) reduces to Eq. (\ref{e:OA-PBE}) for closed-shell systems but improves over Eq. (\ref{e:OA-PBE}) for open shells, as shown by the gray line labeled `OA-PBE' in Fig. \ref{fig:H2p-and-He2p} for H$_2^+$ and in ref. \cite{nafziger2015fragment} for Li$_2^+$. 

Eq. (\ref{e:OA-PBE-II}-\ref{e:dEHnad}), together with the model for $S[\bn]$ of Eq. (\ref{e:overlap}) can even significantly improve the PBE binding energies of He$_2^+$, a challenging molecule for reasons other than delocalization and static-correlation \cite{gruning2001failure}.
The agreement between OA-PBE and CCSD(T) in this case is not quantitative (panel ${\bf b}$ in Fig. \ref{fig:H2p-and-He2p}), but P-DFT calculations indicate that the main source of error belongs to $\Ts^{\rm nad}(R)$, probably due 
to the numerical difficulties associated with finding a pure-state spin-density for such a stretched system \cite{chayes1985density, ullrich2002degeneracy, lammert2006coarse}, leading to a non-zero $T_s^{\rm nad}(R)$ as $R\to\infty$ (see panel ${\bf d}$ of Fig. \ref{fig:H2p-and-He2p}). However,
the results labeled by `kinOA-PBE' in panel \textbf{b} of Fig. \ref{fig:H2p-and-He2p} demonstrate that this error can be suppressed almost entirely by multiplying $\Ts^{\rm nad}$ by $S[\bn]$ and approximating $Q[\bn]$ as $Q[\bn]\approx S[\bn]\Ts^{\rm nad}[\bn]+E_{\rm Hxc, PBE}^{\rm nad,OA}[\bn]$.  

\vspace{0.5cm}
{\bf {\em Summary and Outlook:}} 
 By using molecular spin-densities $\{n_{\uparrow}(\br),n_{\downarrow}(\br)\}$ as the main variables in calculations, XC approximations are hard-pressed to describe the low-density inter-nuclear regions of molecules where correlation effects are relatively more important (compared to kinetic effects), so the XC approximations are in a sense blind to the formation of fragments when bonds are stretched.  Methods including symmetry-breaking \cite{perdew2021interpretations}, self-interaction error corrections \cite{perdew1981self, mori2006many, pederson2014communication, schmidt2016one, yang2017full}, range-separated functionals \cite{savin1995density, iikura2001long, yanai2004new, vydrov2006assessment, chai2008long, baer2010tuned}, double hybrid functionals \cite{grimme2006semiempirical, schwabe2007double}, and scaling correction methods \cite{li2018localized} are all among approaches that have been adopted to overcome such difficulties. Moreover, methods relying on the “on-top” pair density \cite{perdew1995escaping}, complex orbitals \cite{lee2019kohn}, exact strong-interaction limit functionals \cite{malet2012strong}, and fractional-spin localized orbital scaling corrections \cite{su2018describing} can improve the accuracy for systems of the strongly-correlated type. Recognizing that such strongly-correlated systems are often composed of weakly overlapping fragments \cite{georges1992hubbard, kotliar2006electronic},
the central result of our work is that, when {\em fragment} spin-densities are used as the main variables,
the typical delocalization and static-correlation errors of approximate $E\xc[n_{\uparrow},n_{\downarrow}]$ \cite{burke2012perspective, cohen2012challenges, mardirossian2017thirty, becke2014perspective, PPLB82, mori2009discontinuous} can be largely avoided without having to abandon essential symmetries. This alternative strategy rests on maintaining the use of the same approximate $E\xc[\{n_{\a,\uparrow},n_{\a,\downarrow}\}]$ {\em within} the fragments while introducing new inter-fragment approximations for $E\xc^{\rm nad}[\bn]$. The latter is a functional of the set of fragment spin-densities $\bn$ rigorously defined wihin P-DFT \cite{nafziger2014density}.

Two exact constraints satisfied by $E\xc^{\rm nad}[\bn]$ were used in the construction of Eq. (\ref{e:OA-PBE-II}): (1) $E\xc^{\rm nad}[\bn] \to 0$ as $R\to\infty$, where $R$ denotes the separation between fragments; and (2) $E\xc^{\rm nad}[\bn]\to -E\H^{\rm nad}[\bn]$ for single-electron bonds.  Eq. (\ref{e:OA-PBE-II}), and the accompanying model for $S[\bn]$ in Eq. (\ref{e:overlap}) should be seen as initial attempts at approximating these quantities.  Future approximations to $E\xc^{\rm nad}[\bn]$ should incorporate more exact constraints. For example, how could Eq. (\ref{e:OA-PBE}-\ref{e:overlap}) be improved to encompass van der Waals interactions \cite{ALL96,vv10}?

\vspace{0.5cm}
{\bf {\em Methods:}} 
All calculations were done using a P-DFT implementation in \textit{Psi4} \cite{parrish2017psi4}. 
The cc-pVTZ basis set was used for all molecules in this work, except for the cases of H$_{10}$ and Cr$_2$, for which cc-pVDZ was used instead. P-DFT PBE calculations (without the OA) were checked to differ usually by less than $0.1$ Kcal/mol when compared to direct restricted-PBE results from KS-DFT with the same densities.
The Wu-Yang algorithm \cite{wu2003direct} implemented on Gaussian basis sets in \textit{n2v} \cite{shi2022n2v} was used to calculate all $\Ts[\nf]$ components. The OA was performed as a post-PDFT
approximation, i.e. using the fragment densities yielded by P-DFT \cite{nafziger2015fragment}. The exact OA functional $S$ for Cr$_2$ is defined as ${E_{\rm xc, PBE}^\nad}/(E_{\rm exact} - \Ef - \Ts^\nad - \EH^\nad)$ by assuming the error is entirely contained in $\Exc^\nad$.
Convergence of the partition potential $\vp$ was achieved in each case by updating it iteratively according to:
\begin{equation}
    \vp^{k+1} = \vp^k + \lambda(v_{\rm xc, PBE}[\nf] - v_{\rm xc, inv}[\nf])
    \label{eqn:vp_iterative}
\end{equation}
for the $(k+1)^{\rm th}$ step, where $\lambda$ is the step size. $v_{\rm xc, PBE}[\nf]$ is the XC potential for a choice of XC approximation (we use PBE in this article). $v_{\rm xc, inv}$ is the effective 
patition potential calculated from inversion, as explained by Eq. (\ref{eqn:entire_inversion}) below.
The derivation of Eq. (\ref{eqn:vp_iterative})
is outlined next omitting spin indices for simplicity. Start from the definition of $\vp$ \cite{nafziger2014density} as:
$    \vp = \frac{\delta \Ep}{\delta \na}$,

where $\na$ is the density for fragment $\alpha$. At convergerence, the same $\vp$ is shared by all fragments and is independent of the fragment index $\alpha$.
By separating $\Ep$ as suggested in the main text, $v_p$ is decomposed as:
\begin{equation}
    \vp(\vr) 
    = \frac{\delta \Ts[\nf]}{\delta \nf(\vr)} - \frac{\delta \Ts[\na]}{\delta \na(\vr)} 
    + v(\vr)-v_{\alpha}(\vr) 
    + \vH[\nf](\vr)-\vH[\na](\vr)
    + v_{\rm xc, PBE}[\nf](\vr)-v_{\rm xc, PBE}[\na](\vr).
    \label{eqn:vp_nad}
\end{equation}
Given that stationary condition for the fragments at each step $k$:
\begin{equation}
    \frac{\delta \Ts[\na]}{\delta \na(\vr)} + v_{\alpha}(\vr) + \vH[\na](\vr) + v_{\rm xc, PBE}[\na](\vr) + \vp = \mu_\alpha,
    \label{eqn:frag_stationary}
\end{equation}
as well as for the entire system through inversion:
\begin{equation}
    \frac{\delta \Ts[\nf]}{\delta \nf(\vr)} + v(\vr) + \vH[\nf](\vr) + v_{\rm xc, inv}[\nf](\vr) = \mu,
    \label{eqn:entire_inversion}
\end{equation}
Eq. (\ref{eqn:vp_iterative}) follows by omitting the chemical potentials since $\mu$ provides no energy contribution to the total energy.

\bibliography{Yuming}
\end{document}